\documentclass[twocolumn]{revtex4}
\usepackage{graphicx}
\usepackage[]{epsfig}
\usepackage{epstopdf}

\newcommand{\non}{\nonumber}

\renewcommand{\b}{\beta}

\newcommand{\e}{\epsilon}

\newcommand{\s}{\sigma}

\newcommand{\comment}[1]{}
\newcommand{\beq}{\begin{equation}}
\newcommand{\eeq}{\end{equation}}
\newcommand{\beqd}{\begin{displaymath}}
\newcommand{\eeqd}{\end{displaymath}}
\newcommand{\beqa}{\begin{eqnarray}}
\newcommand{\eeqa}{\end{eqnarray}}
\newcommand{\Tr}{{\rm Tr}\,}

\def\HH{{\cal H}}
\def\OO{{\cal O}}

 \let\b=\beta    \let\d=\delta \let\e=\varepsilon

\let\s=\sigma     
   
   \let\L=\Lambda 
         
  \let\r=\rho

 \def\HH{{\cal H}}
 
  \def\OO{{\cal O}}
 \def\SS{{\cal S}}

\def\to{\rightarrow}

\begin{document}

\title{On the critical slowing down exponents of mode coupling theory}

\author{F. Caltagirone$^{1}$,   U. Ferrari$^{1,2}$, L. Leuzzi$^{1,2}$,
G. Parisi$^{1,2,3}$, F. Ricci-Tersenghi$^{1,2,3}$ and T. Rizzo$^{1,2}$}

\affiliation{$^1$ Dip. Fisica, Universit\`a {\em La Sapienza}, Piazzale
  A. Moro 2, I-00185, Rome, Italy \\ $^2$ IPCF-CNR, UOS Roma {\em
    Kerberos}, Universit\`a {\em La Sapienza}, P. le A. Moro 2, I-00185,
  Rome, Italy \\ $^3$ INFN, Piazzale A. Moro 2, 00185, Rome, Italy}

\begin{abstract}
A method is provided to compute the parameter exponent $\lambda$
yielding the dynamic exponents of critical slowing down in mode
coupling theory.  It is independent from the dynamic approach and
based on the formulation of an effective static field theory.
Expressions of $\lambda$ in terms of third order coefficients of the
action expansion or, equivalently, in term of six point cumulants are
provided.  Applications are reported to a number of mean-field models:
with hard and soft variables and both fully-connected and dilute
interactions.  Comparisons with existing
results for Potts glass model, ROM, hard and soft-spin
Sherrington-Kirkpatrick and p-spin models are presented.
\end{abstract} 

\maketitle

In the framework of glassy mean-field (MF) models with quenched or
built-in disorder, it is well known that the dynamics of models whose
glassy phase is consistently described by a Replica Symmetry Breaking
(RSB) solution with a finite number of breaking's displays critical
slowing down and a dynamic transition \cite{Gardner85,
  Kirkpatrick87c,Crisanti93,Marinari94a}. The equations governing the
relaxation dynamics down to the dynamic critical temperature $T_d$ are
those pertaining to the schematic mode-coupling-theory (MCT) developed
in the context of supercooled liquids
\cite{Bengtzelius84,Goetze84,Goetze89, Goetze09,Bouchaud96}.  The
``guide observable" is the correlation function $C(t)$, i.e., the
overlap between a given initial equilibrium configuration of the
dynamics and the configuration at time $t$. In the discontinuous one
step RSB case near $T_d$ the relaxation of $C(t)$ is a two step
process in which the system spends a large amount of time,
algebraically diverging as $T\to T_d$, around a plateau value $q_{EA}$.  Models
with a discontinuous dynamic transition include, e.g., the spin-glass
(SG) $p$-spin model with either spherical or Ising spins, the Potts
glass model and the Random Orthogonal Model (ROM).
\indent
MCT predicts that two exponents control the whole dynamics. Exponent
$a$ governs the early $\beta$ regime as $C(t)$ approaches the plateau
value with a power law $C(t)\approx q_{EA}+c_a/t^a $, while exponent
$b$ identifies the early $\alpha$ regime as $C(t)=q_{EA}-c_b t^b$,
when the system starts relaxing towards equilibrium \cite{Goetze09}.
An important prediction of MCT is the following relationship between
the decay exponents and their mutual relationship to the so-called
``exponent parameter" $\lambda$:
 \beq {\Gamma^2(1-a) \over \Gamma(1-2
  a)}={\Gamma^2(1+b) \over \Gamma(1+2 b)}=\lambda
\label{uno}
\eeq
\indent In case of a continuous transition, there is no dynamic arrest
and no $b$ exponent is defined. Well known instances are, e.g., the
paramagnet to full-RSB SG transition along the de Almeida Thouless (dAT)
line in mean-field SG models, either fully connected
\cite{Sompolinsky81,Sompolinsky82} or on random graphs
\cite{Carlson90}, as well as the SG transition in Potts models with $p
\leq 4$, both fully connected \cite{Gross85} and on the Bethe lattice
of any connectivity \cite{Mulet02}, and in the $p$-spin spherical
model with large external magnetic field \cite{Crisanti93}.
\indent
Even though Eq. (\ref{uno}) is usually assumed as the correct
relationship between exponents $a$ and $b$, the exponent parameter
$\lambda$, apart from schematic MCT cases, is simply considered a
tunable parameter, without a specific connection to any physical
observable. Here we show how to unveil this
connection in full generality working within a ``static-driven"
effective theory of dynamics: we will put forward an independent
formulation of $\lambda$ and apply it to some paradigmatic SG models.
\indent In general, the analytic treatment of the dynamics is more
complicated than the statics and only a few models have been studied
so far: the soft-spin Sherrington-Kirkpatrick (SK) model
\cite{Sompolinsky81,Sompolinsky82}, schematic MCT's \cite{Goetze84},
soft-spin $p$-spin and Potts glass models, for which the connection
with MCT was first identified \cite{Kirkpatrick87c}. This prompted to
consider the spherical $p$-spin SG \cite{Crisanti92,Crisanti93} in all
details as a MF structural glass, cf., e.g., Ref. \cite{Cavagna05},
even in the off-equilibrium regime below $T_d$ \cite{Cugliandolo93}.
In these cases dynamics is explicitly solved and $\lambda$
exactly computed. In particular, one finds that {\it it is not
  universal} and depends on model and  external parameters.
On the other hand, its computation becomes difficult when we consider
more complicated MF systems and finite-dimensional ones.


{\em Static definition of the exponent parameter $\lambda$.} $\qquad$
Similarly to the static transition also the dynamic one can be located
as the critical point of an appropriate replicated Gibbs free energy
$\Gamma$ \cite{Monasson95,Franz98b, Franz11b}. This is a
function of the replicated dynamic variables (e.g. spins $\s$), Legendre
transform of the replicated free energy $\Phi$
\begin{equation}
\Gamma[\delta Q_{ab}]= \Phi[\Lambda_{ab}] + \sum_{ab}\Lambda_{ab}\delta Q_{ab},  \quad \frac{\partial \Phi}{\partial \Lambda}=\delta Q_{ab} 
\end{equation}
 with respect to a conjugated field $\Lambda$, function of
 the overlap matrix. We used the short-hand $\delta Q_{ab} = \s_a
 \s_b-q$; $q=\langle \s_a \s_b\rangle$ being the Edwards Anderson
 parameter, that is the value of the overlap matrix elements for the
 replica symmetric (RS) solution.  Average $\langle \ldots
 \rangle$ is performed over the proper replicated ensemble
 \cite{Mezard87}
\footnote{This is equivalent to ${\overline{\langle \ldots
      \rangle_J}}$, where brackets and overline denote, respectively,
  ensemble (at fixed disorder $J$) and disorder average.  When
  disorder is self-induced this is equivalent to the overall
   thermal average including nested average over pinning
  and pinned variables \cite{Monasson95,Franz95a,Franz97,Cammarota11}.}.  
Expanded around the RS critical point $\Gamma$ reads:
\begin{eqnarray}
\Gamma(\delta Q)&=&{1 \over 2}\sum_{(ab),(cd)}\delta Q_{ab} M_{ab,cd} \delta Q_{cd}
\label{f:Gibbs_fe}
\\
\nonumber
&&-{w_1 \over 6}\Tr \delta Q^3-{w_2 \over 6}\sum_{ab}\delta Q_{ab}^3
\end{eqnarray}
with $a,b,c,d=1, \ldots, n$.  We have retained only two of all cubic
coefficients for they are the relevant ones at criticality
\cite{Temesvari02b}. The case $n=0$ is relevant for the continuous
transition \cite{Edwards75,Sherrington75,Parisi79}, the case $n=1$ for
the dynamic discontinuous transition
\cite{Monasson95,Franz98b,Crisanti08,Franz11b}.  Our main
result is: 
\beq \lambda={w_2 \over w_1}
\label{result}
\eeq 
We note that this ratio also yields the breaking point $x$ in the
case of continuous transition from RS to full-RSB \cite{Gross85}. The
above result can be obtained in the context of the supersymmetric
formulation of the dynamics \cite{Kurchan91} that
is a very convenient way of seeing the connection between equilibrium
dynamics and the static replica treatment. Details of the derivation
will be given elsewhere \cite{Rizzo11}.

Eq. (\ref{result}) holds in full generality above the upper critical
dimension. In general, we do not have an analytic expression of the
Gibbs free energy, e.g., for MF models defined on finite-connectivity
random graphs. However, $\Gamma$ is defined as the Legendre transform
of $\Phi$ and, therefore, its proper vertices can be associated to
cumulants of the replicated order parameter.  We, thus, face the
problem of computing $\Gamma$ from $\Phi$ in presence of fields
$\Lambda_{ab}$ coupled to $1/N\sum_i s_i^a s_i^b$.  The free energy
$\Phi$ needs to be computed at the third order, i.e., dealing
with eight RS cumulants $\omega_{1,\ldots,8}$
\cite{Temesvari02b}.  The coefficients $w_1$ and $w_2$ of the Gibbs
potential, cf. Eq. (\ref{f:Gibbs_fe}), however, are expressed as
\beq
w_1=r^3 \omega_1 \ \ \ w_2=r^3 \omega_2 \eeq 
where $r$ is the inverse of the SG susceptibility: 
\beq r=
\left[ {1 \over N}\sum_{ij}\overline{\langle s_i s_j\rangle_c^2}
  \right]^{-1} \eeq 
i.e., the replicon, and $\omega_1$ and $\omega_2$ can be written in
terms of six spin correlations (details elsewhere \cite{Rizzo11}):
\beq \omega_1={1 \over N}
\sum_{ijk}\overline{\langle s_i s_j\rangle_c \langle s_j s_k\rangle_c
  \langle s_k s_i\rangle_c}
\label{omega1}
\eeq
\beq
\omega_2={1 \over 2 N} \sum_{ijk}\overline{\langle s_i s_j s_k\rangle_c^2 }
\label{omega2}
\eeq
where ${\overline{ {\phantom I}\ldots {\phantom I}}}$ means average
over the disorder and $\langle \ldots \rangle_c$ denotes the connected
thermal average.  In case of discontinuous transition it is implicit
that different thermal averages in the above expression are all
computed within the same state, selected by the initial condition.  We
underline that while vertices $w_1$ and $w_2$ remain finite at $T_d$,
the corresponding $\omega$ cumulants diverge as $r^{-3}$.

{\em Method validation and $\lambda$ computation.} $\quad$ Applying
Eq. (\ref{result}) and computing $w_1$ and $w_2$ within the above
mentioned static approach, we have tested our prediction on
 various models.  In those few cases where $\lambda$ is
exactly known from the dynamics
\cite{Sompolinsky82,Crisanti93,Paoluzzi11b} we verify that its formal
expressions are identical.  In more complicated systems the dynamic
MCT phenomenology has been studied numerically and estimates of the
exponents are available in the literature. These are instances
in which Eq. \ref{result} yields actual analytic prediction for $\lambda$.

{\em Fully connected models.}  $\quad$In the following, we consider a
family of models with a Hamiltonian of the kind:
\beqa \HH = -\sum_{i<j} J_{ij} \s_i \s_j - \sum_{p=1}^\infty\!
\sqrt{\!\frac{R^{(p)}}{p!} }\!\sum_{i_1< \dots < i_p}\!
\!\!\!K^{p}_{i_1, \dots, i_p} \s_{i_1} \cdots \s_{i_p} 
\nonumber
\eeqa
where $\s_i$ are $N$ Ising spins, or soft/spherical ones. The $2$-body
interaction matrix is constructed as $J=\OO^{T}{\cal E}\OO$, where
$\OO$ is a random $O(N)$ matrix chosen with the rotational invariant
Haar measure, cf., e.g., Ref. \cite{Edelman05} and ${\cal E}$ is a diagonal
matrix with elements independently chosen from a distribution $\r(\e)$
\cite{Cherrier03b}. In order to ensure the existence of the
thermodynamic limit, the support of $\r(\e)$ must be finite and
independent of $N$.  The $p$-body interactions $K^{(p)}$ are
i.i.d. Gaussian variables with zero mean and variance $p!/N^{p-1}$ and
$R^{(p)}=d^p R(x)/dx^p (x)|_{x=0}$
for some real vauled function $R(x)$.  For the MF Ising SG
[\onlinecite{Sherrington75,Gardner85,Marinari94a}], as well as for
 spherical SG's \cite{Crisanti92,Crisanti04,Crisanti07b},
the general form of the replicated free energy is:
\beqa
-n\b \Phi&=&\mathrm{extr}_{Q,\L}  \SS[Q,\L] 
\\
 \SS[Q,\L] & = &\frac{1}{2} \Tr \,G(\b Q) + \frac{\beta^2}{2}
\sum_{ab} \,R( Q_{ab}) 
\label{action}
 \\ 
\non
&& - \frac{1}{2} \Tr \, Q \L + \ln  \Tr_{\{\s\}} \, {\cal W}[\L;\{\s\}]
\\
{\cal W}[\L;\{\s\}]&=&
\exp\Biggl\{
\frac{1}{2} \sum_{a,b} \L_{ab} \,\,
  \s_a \s_b\Biggr\} 
\eeqa
\noindent
where $G:\, M_{n\times n} \to M_{n\times n}$ is a function in the
space of $n\times n$ matrices, formally defined through its power
series around zero. Its form depends on the choice of the eigenvalue
distribution $\r(\e)$ of the ${\cal E}$ matrix.

Given this effective action $\SS[Q,\L]$, the saddle point equations in
$\Lambda$ and $Q$ respectively read
\beqa
Q_{ab} &=&\langle \s_a \s_b \rangle_{\cal W} 
\label{saddle_romQ}\\
\L_{ab} &=& \b [G'(\b Q)]_{ab} + \b^2 R'(Q_{ab}) 
\non
\eeqa
that, in the RS Ansatz, become
\beqa 
q &= &\langle m^2(z)  \rangle_z ; \qquad m(z)=\langle\s\rangle_{\s}\\
\L &= &\frac{\b}{n} [ G'(\b(1+(n-1)q)) -  G'(\b(1-q))] 
+ \b^2 R'(q)
\nonumber
\eeqa
where the weights over which $\langle \ldots \rangle_{z}$ and
$\langle\ldots \rangle_{\s}$ are performed depend on $\s$ being Ising,
spherical or soft. For the cases of interest here, weights are 
proportional to the following distributions:
\begin{eqnarray}
\nonumber
\begin{array}{lc|c}
&\langle\ldots\rangle_{\s} & \langle\ldots\rangle_{z}\\
{\rm Ising}&\ \ e^{z\s}[\delta(\s+1)+\d(\s-1)]&\ \ e^{-z^2/(2\L)}\cosh^n(z)\\
\hline
{\rm Spher.}&\ \ e^{z\s}\exp\{-\s^2/[2(1-q)]\}&\ \ e^{-z^2/(2\L)}e^{n(1-q)z^2/2}
\end{array}
\nonumber 
\end{eqnarray}
\noindent implying $m(z)=\tanh(z)$ for Ising and $(1-q)z$ for
spherical spins.  To compute the values of $w_{1,2}$,
cf. Eq. (\ref{f:Gibbs_fe}), one has to expand Eq. (\ref{action}) to
third order around the saddle point value.  Considering the second
order term and imposing that the Hessian determinant vanishes
(criticality condition) we obtain
\beq
\langle\langle (\s-m)^2\rangle_{\s} ^2\rangle_z = \left[\b^2 G''(\b
  (1-q)) + \b^2 R''(q)\right]^{-1}
\label{criticality}
\eeq

Using the above condition and expanding Eq. (\ref{saddle_romQ}) to
second order, we derive the general expressions of $w$'s for
fully-connected systems:
\begin{eqnarray}
\label{f:w1}
\frac{3!}{\beta^2} w_1\!\!&=&\!\! \frac{\b}{2} G'''(\b (1-q))+
A(q) \langle \langle(\s-m)^2\rangle_{\s}^3 \rangle_z 
\\ 
\frac{3!}{\beta^2} w_2\!\!&=&\!\! \frac{R'''(q)}{2}+2 A(q) \langle \langle (\s-m)^3\rangle_{\s}^2 \rangle_z
    \label{f:w2}
    \\
    \non
    A(q)&\equiv&\beta^4 [G''(\b (1-q)) +    R''(q)]^3
    \end{eqnarray}
 holding both for $n=0$ and $n=1$. They can be used to compute
 $\lambda$ in different cases. We now exemplify a few.

For the {\underline{SK model}}, $R(x)=x$, $G(x)=x^2/2$,
$\rho(\e)=\sqrt{4-\e^2}/(2\pi)$, Sompolinsky's result for Ising spins
along the dAT line is recovered \footnote{To be precise, at difference
  with the original work we have a Gaussian field $h_i$, with
  ${\overline h_i}=0$ and ${\overline {h_i h_j}}=\delta_{ij}h^2$, but
  the two cases can be shown to be equivalent \cite{Parisi98}.},
cf. Eq. 2.61 of Ref.  \cite{Sompolinsky82}.
 
For the {\underline{ ROM model}} \cite{Marinari94a, 
  Cherrier03b,Parisi95b}, one has $R(x)=0$,
\begin{eqnarray}
2G(x)&=& v_\alpha(x)-1+2(\alpha-1)\ln \frac{v_\alpha(x)+2x+2\alpha-1}{2\alpha}
\non
\\
\non
&&-\ln \frac{v_\alpha(x)+1+2x(2\alpha-1)}{2}
\non
\\
\non
v_\alpha (x)&\equiv&\sqrt{1+4x(2\alpha-1+x)} 
\end{eqnarray}
$\rho(\e)=\alpha\delta(\e-1)+(1-\alpha)\delta(\e+1)$, and the
transition is dynamic.  MCT dynamics in the ergodic phase has been
numerically studied in Ref. \cite{Sarlat09} for $\alpha=13/32$, where
strong finite-size effects are observed and two different estimates
for the exponent provided: $b=.62$, from the fit of the von Schweidler
law, while $b=.75$, from the fit of the equilibrium $\alpha$
relaxation time vs. temperature. Our ``static-driven" computation
yields $b=.628$ ($\lambda=.7077$), allowing for a validation of the
first numerical estimate.

For {\underline{ Ising $p$-spin}}, $G=0$ and $R(x)=x^p/2$, $\lambda$ values are
 in Tab. I.  For $p\to 2$ we retrieve the result of Ref.
\cite{Kirkpatrick87c}.
 \begin{table}
\label{tab:Isip}
 \begin{tabular}{c|c|cccccccccc} 
 $p$ & $\to 2$ & $2.05$ &$2.2$&$2.5$&$3$&$4$&$5$&$6$&$7$&$8$&$9$ \\
 \hline
 $\lambda$ &$.5$&$.556$&$.652$&$.719$&$.743$&$.746$& $.743$&$.739$& $.736$&$.733$&$.731$ \\
\hline $a$ &$.395$& $.379$&$.346$&$.32$&$.308$&$.306$&$.308$&$.31$&$.311$&$.313$&$.314$
 \end{tabular}
 \caption{Dynamic exponents in the Ising $p$ spin model.}
  \vskip -.3cm \end{table}
\\
\indent
  For {\underline {spherical models}}, $G=0$, $R(x)=\sum_r x^r/2$, the
  exact analytic form $\lambda=\Lambda''(q_d) (1-q_d)^3/2$ is
  retrieved \cite{Crisanti93,Paoluzzi11b,Franz11c}, equivalent to the schematic
  MCT prediction for very long times at criticality, i.e., for
  $C(t)=q_d$ \cite{Goetze89,Bouchaud96}, ${\L}(C(t))$ being the MCT
  equations memory kernel.
\\
\indent
Yet another example is the {\underline{ Potts glass}}:
\beq
{\cal H}=-\sum_{ij}J_{ij}  \left(p~ \delta_{\sigma_i\sigma_j}-1\right)
\eeq
The model has a discontinous glass transition for $p>4$
\cite{Gross85}.  Brangian {\em et al.}  studied the Potts glass with
$p=10$ by means of Monte-Carlo numerical simulations
\cite{Brangian02}.  Approaching the dynamic transition, finite-size
effects turn out to be large, implying that the {\it plateau} is
almost invisible also for very large sizes: this makes the numerical
estimation of exponents very difficult.  Their interpolation yields
$a=.33\pm .04$.  For $p=10$, from the expansion of $\Gamma[\delta Q]$
around $q=q_d$, we obtain the exact values $\lambda=.8053$ and $a=.2759$
\cite{Calta11}, compatible with, though not extremely near to, the
numerical estimate.

\begin{figure}[t!]
\begin{center}
\epsfig{file=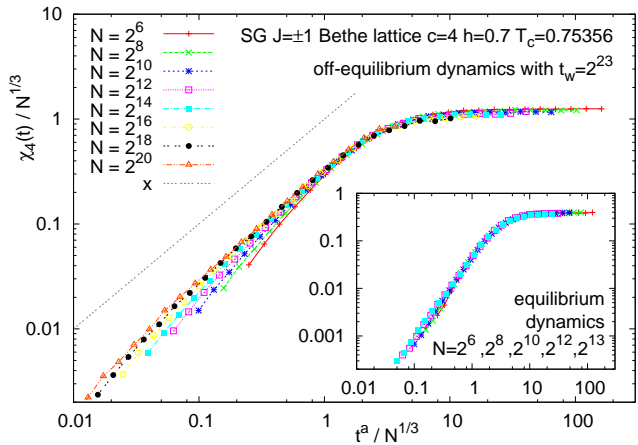,width=.5\textwidth}
\caption{Rescaled $\chi_4(t)$ for the Viana-Bray model with
  $c=4$, $h=.7$ and $T=.73536(1)$, along the dAT line with  
$a=a_{\rm th}=0.406$. No fitting
  interpolation is performed.}
\label{CDS}
\end{center}
\end{figure}

{\em Models on diluted random graphs.} $\quad$ To study glassy models
on random graphs we set up an apart technical method to analytically
compute dynamic exponents.  To frame the results, we first recall that
there is no single instance of these class of models whose dynamics
has been solved explicitly. Eq. (\ref{result}) would allow to bypass
the problem provided one had the replicated action $\Gamma$.
Unfortunately, the action is hard to compute in diluted models, thus
preventing $w$'s derivation.  We can, however, explicitly compute the
related cumulants $\omega_{1,2}$ at criticality, cf.
Eqs. (\ref{omega1})-(\ref{omega2}).
Details of the derivation will be given elsewhere \cite{Calta11},
while here we present a validation of the results for the
Viana-Bray model \cite{Viana85} in a field, displaying a continuous
transition along the dAT line.
\\ \indent Let us consider a Bethe lattice with connectivity $c=4$, in
a field $h=.7$ at the corresponding critical temperature
$T_c(h)=0.73536(1)$ \cite{Parisi11}. The cumulants ratio yields
$\lambda=0.461$.  Our analityc prediction for the $\beta$ decay
exponent is, thus, $a_{\rm th}=.406$.  We, then, numerically study the
system by means of both equilibrium and off-equilibrium Monte Carlo
simulations.  At equilibrium sizes $2^8 \le N \le 2^{13}$ have been
probed.  Starting from an equilibrated configuration we measure the
decay of $C(t)\equiv \sum_i s_i(0)s_i(t)/N$ with time. For each sample
$6$ replicas are simulated and $16$ uncorrelated measurements taken
for each replica. Number of samples: from $1280$ ($N=2^{13}$) to
$5120$ ($N=2^6$).  Off-equilibrium, the two time $C(t,t_w)$ is
measured. For very long $t_w$, $C(t,t_w)$ tends to the equilibrium
$C(t)$.  For each sample we take one measure (at the longest simulated
$t_w$). The number of samples goes from $1280$ ($L=2^{20}$) to $51200$
($L=2^6$).  Due to finite size effects, $C(t)$ (at and off-
equilibrium) displays a power law behavior only for times smaller than
a time scale $t^*(N)$ that diverges with $N$. This makes the
estimation of the exponent hard. Therefore, rather than $C(t)$, we
probe its thermal and disorder fluctuations,
\begin{equation}
\chi_4(t)= N \left[\overline{\langle
    C(t)^2\rangle}-\overline{\langle C(t) \rangle}^2\right]
\label{f:chi4}
\end{equation}
remaning finite for large $N$ at finite $t$.  Using finite-time
scaling arguments it can be argued that (i) $\chi_4(t)$
diverges as $t^a$ at large times on a scale smaller that $t^*(N)$,
(ii) the critical region diverges as $t^*(N)=N^{1/(3 a)}$, (iii) on
times scales larger than the critical region the fluctuations scale as
$N^{1/3}$ \cite{Parisi93a}.  As a consequence, if $a$ has the correct
value, the rescaled dynamic critical $\chi_4(t)/N^{1/3}$ vs.
$t/N^{1/(3a)}$, should be size independent. We plot it in
Fig. \ref{CDS} with $a=a_{\rm th}=0.406$.  Collapse appears excellent
both at and off-equilibrium.
\\ \indent Concluding, we have introduced a ``static-driven'' method
to obtain, by means of a replica field theory, the dynamic exponents
of the critical slowing down. The method allows to determine the MCT
parameter exponent $\lambda$ as the ratio of coefficients of third
order terms of the Gibbs free energy action expanded around the
critical point, cf. e.g., Eqs. (\ref{f:w1}-\ref{f:w2}).  Equivalently,
$\lambda$ is shown to be equal to the ratio between six points
cumulants of a theory whose action is the Legendre transformed of the
Gibbs free energy, cf. Eqs. (\ref{omega1}-\ref{omega2}). Indeed, {\em
  the dynamical exponents can be associated to the ratio between two
  physical observables computed within a static framework}. We
verified the method's prediction in various MF models, both on
fully-connected and diluted graphs, successfully comparing with
previous analytical and numerical results.  The method can be exported
to any glass models whose Gibbs action is computable or whose six
point cumulants can be estimated.

{\em Acknowledgements.} ~~ We thank A. Crisanti, S. Franz and E.
Zaccarelli for useful discussions.

\end{document}